\documentclass[onecolumn,showpacs,amsmath,amssymb,amsfonts,a4paper,aps,prd,10pt]{revtex4}
\usepackage{graphicx}
\usepackage{epstopdf}
\newcommand{\nc}{\newcommand}
\nc{\ba}{\begin{eqnarray}} \nc{\ea}{\end{eqnarray}}
\newcommand\be{\begin{equation}}
\newcommand\ee{\end{equation}}
\nc{\D}{\overline{\mbox{D3}}}

\nc{\ga}{\gamma} \nc{\tnu}{\tilde{\nu}} \nc{\tmu}{\tilde{\mu}}

\nc{\x}{{\bf{x}}}
\newcommand{\f}[2]{\frac{#1}{#2}}
\newcommand{\mc}[1]{\mathcal{#1}}
\newcommand{\tg}{\tilde{g}}
\newcommand{\tf}{\tilde{f}}
\renewcommand{\th}{\tilde{h}}
\input{epsf.sty}
\begin{document}

\title{Multi-Metric Gravity via Massive Gravity}
\author{Nima Khosravi$^{1,2}$}
\email{nima@ipm.ir}
\author{Nafiseh Rahmanpour$^{3}$}
\email{nafisehrahmanpour@gmail.com}
\author{Hamid Reza Sepangi$^{3}$}
\email{hr-sepangi@sbu.ac.ir}
\author{Shahab Shahidi$^{3}$}
\email{s_shahidi@sbu.ac.ir}

\affiliation{$^1$Cosmology Group, African Institute for Mathematical
Sciences (AIMS), Muizenberg 7945, South Africa}

\affiliation{$^2$Abdus Salam International Centre for Theoretical Physics,
Strada Costiera 11, 34151, Trieste, Italy}

\affiliation{$^3$Department of Physics, Shahid Beheshti University,
G. C., Evin, Tehran 19839, Iran}

\begin{abstract}
A generalization to the theory of massive gravity is presented which
includes three dynamical metrics. It is shown that at the linear
level, the theory predicts a massless spin-2 field which is
decoupled from the other two gravitons which are massive and
interacting. In this regime the matter should naturally couple to massless
gravitons which introduce a preferred metric that is the average of the primary metrics. The
cosmological solution of the theory shows the de-Sitter
behavior with a function of mass as its cosmological constant.
Surprisingly, it lacks any non-trivial solution when one of the
metrics is taken to be Minkowskian and seems to enhance the
predictions which suggest that there is no homogeneous,
isotropic and flat solution for the standard massive cosmology.

\end{abstract}
 \maketitle
 \tableofcontents

\section{Introduction and Summary}
The general theory of relativity (GR) has been reigning supreme
since its introduction by Einstein nearly a hundred years ago. The
underlying principles upon which GR is founded, namely the
principles of equivalence and general covariance, have given the
theory, in spite of its inevitable shortcomings, an unprecedented
breath and depth, enabling it to predict, solve and answer questions
which seemed beyond the possibilities of any scientific endeavor
over the past decades. That said, new observations and discoveries
in the recent past seem to be beyond the predictive powers of the
standard GR. The discovery of the accelerated expansion of the
universe relating to dark energy and galaxy rotations curves
relating to dark matter are some of the examples that necessitate
modifications to the standard GR. Among many proposals to modify GR,
Massive Gravity (MG), whose roots go back to the work of Fierz and
Pauli  \cite{fierz} in 1939 has been gaining momentum in recent
years after being brought back from  oblivion in the 1970's where
renewed interest in quantum field theory became pervasive.  It is an
effective field theory of gravity which, as the name suggests, is an
extension of gravity with a massive spin-2 graviton. Its linear
version coupled to a source was studied by van Dam, Veltman and
Zakharov \cite{van} who discovered that it makes predictions
different from the linear GR even in the limit of the zero mass.
This is known today as the vDVZ discontinuity. This problem is
traced back to the degrees of freedom entering the theory,  5 for a
massive and 2 for a massless spin-2 graviton. Attempts to remedy
this shortcoming is suggested in \cite{vain}. However, Boulware and
Deser \cite{deser} demonstrated that the theory suffers from the
existence of ghosts, when non-linear self-interaction terms added. A
fix was suggested in \cite{berg} where it was demonstrated that the
resulting theory is unitary and ghost free. A comprehensive and well
written review on this subject can be found in \cite{kurt}.

Recently, it has been proposed in \cite{Hassan-Rosen-bimetric} that
massive gravity in a ghost free form can be written in a bimetric
language. Both of these topics separately have had their supporters
and followers. Bimetric theories are theories with two dynamical
spin-2 fields. However, it has been shown that there are no
consistent theories of interacting massless spin-2 fields. In
general these bimetric theories suffer form having Boulware-Deser
ghosts, and so are unstable \cite{deser}. The bimetric theory was
first proposed by N. Rosen in \cite{N-Rosen-bimetricI} to give a
tensor character to quantities like gravitational energy-momentum
density pseudo-tensor. In this theory the tensor $g_{\mu\nu}$ is
regarded as a gravitational field that has no direct connection to
geometry and so geometrization of gravity is given up
\cite{N-Rosen-bimetricII}. Subsequently a large amount of work has
been done to address cosmological problems in this context
\cite{Drummond,Mond}.

Contrary to N. Rosen in \cite{N-Rosen-bimetricI} and
\cite{N-Rosen-bimetricII}, some of these works have kept the
geometrical viewpoint of General Relativity. Milgrom has used two
metrics for constructing relativistic formulation of MOND gravity
\cite{Mond} which naturally leads to MOND and dark energy effects. In
this BIMOND theory, matter lives in the space-time described by one
of the metrics which couples to another metric through an
interaction term. Drummond in \cite{Drummond} introduced two
vierbein bundles into the space-time manifold so that each bundle
supports its own metric. One of this metrics is associated with
matter and the other with gravity. This theory is a kind of modified
theory of gravity that has the flexibility to permit the
introduction of a length scale to explain the observed rotation
curves of galaxies. On the other hand, in the last two years, the
theory of massive gravity has opened a new era of research after the
work of de Rham, Gabadadze and Tolley \cite{drgt} where a new
interaction term for a non-linear massive gravity was constructed,
leading to a theory which becomes ghost free. The cosmology of such
a model is considered and shown to be devoid of the flat FRW
solutions \cite{cos}. However it has recently been shown that the
open FRW solution does exist \cite{muko}.

In ordinary formalism of massive theories of gravity an additional
reference metric $f_{\mu\nu}$ is required. It has been shown
recently by S. F. Hassan and R. A. Rosen
\cite{Hassan-Rosen-bimetric} that if we use the interaction term
constructed in  massive gravity theories, resulting in a ghost free
massive theory, one will arrive at a bimetric theory of gravity
which is free of the Boulware-Deser ghost. This model is important
in the context of bimetric theories due to the form of definition of
the potential term. The cosmology of such a bimetric theory is studied
in \cite{volkov,hassan}.

In this work we generalize the massive gravity scenarios to include
three dynamical metrics. It will be clear that this trimetric
formalism is very similar to $N$-metric models, though it has new
predictive powers as in bimetric models. Naturally, by introducing
$N$ metrics, $N$ gravitons are expected in its linear regime. This
fact is similar to the Kaluza-Klein modes. In this context
the existence of $N$ gravitons and the interaction between them are
considered to address an effective field theory for gravity
\cite{nima-a-h-1,nima-a-h-2}. In \cite{nima-a-h-1} there
is an interesting interpretation of massive gravitons using the
Fierz-Pauli mass in that the authors assume to have $N$ sites which are related by
interaction through the mass terms. This interpretation can also be used
in our assumption of $N$ metrics, not in the Fierz-Pauli format but
in the new ghost-free massive gravity formalism.

As mentioned above, at the linear level,  we arrive at a theory
which describes one massless spin-2 field which is decoupled from
two interacting massive spin-2 fields. In addition, the cosmological
solution of this model will be presented. This model, similar to
the very recent results in bimetric cosmology \cite{volkov,hassan}, has some
de-Sitter solution with a cosmological constant which is a function
of the masses. It is interesting that in this model which is more
extensive than ordinary massive cosmology \cite{cos}, there is no
non-trivial cosmological solution if one assumes one of the metrics
to be Minkowskian. So it seems that the result of \cite{cos} can be
generalized to: ``there is no non-trivial solution for massive
cosmology (even in its multi-metric representation) by assuming one
Minkowski metric".

\section{The Model}\label{sec2}
We start with a review of the non-linear massive gravity action
presented in \cite{drgt}. The non-linear ghost-free action which
reduces to the Fierz-Pauli action at the linear limit is
\begin{align}\label{eq201}
S_{massive}=-M_{Pl}^2\int
d^4x\sqrt{-g}\left[R-2m^2\sum_{m=0}^4\beta_n
e_n\left(\sqrt{g^{-1}f}\right)\right]
\end{align}
where the square root matrix is defined as $\sqrt{A}\sqrt{A}=A$ for
a general metric $A$. The $e_k(\sqrt{g^{-1}f})$ are 5 elementary
polynomials of the eigenvalues $\lambda_n$ of the matrix
$\sqrt{g^{-1}f}$. They are explicitly written as
\begin{align}\label{eq202}
&e_0(\sqrt{g^{-1}f})=1\nonumber\\
&e_1(\sqrt{g^{-1}f})=\sum_{i=1}^4\lambda_i\nonumber\\
&e_2(\sqrt{g^{-1}f})=\sum_{i<j}\lambda_i\lambda_j\nonumber\\
&e_3(\sqrt{g^{-1}f})=\sum_{i<j<k}\lambda_i\lambda_j\lambda_k\nonumber\\
&e_4(\sqrt{g^{-1}f})=\prod_{i=1}^4\lambda_i=det\sqrt{g^{-1}f}
\end{align}
This implies that the highest order term in the interaction part of
the action is simply a determinant of the background metric
$f_{\mu\nu}$. We also note that one can write the polynomials
$e_k(\sqrt{g^{-1}f})$ in terms of trace of the matrix
$\sqrt{g^{-1}f}$. One  observes that the potential term in the
action is symmetric under the transformation
\cite{Hassan-Rosen-bimetric}
\begin{align}\label{eq203}
f\leftrightarrow g,\qquad \beta_n \leftrightarrow \beta_{4-n},
\end{align}
so one may consider the above interaction term to be the interaction
for the metric $g_{\mu\nu}$ with the background metric $f_{\mu\nu}$
or the interaction for the metric $f_{\mu\nu}$. These observations
led S. F. Hassan and R. A. Rosen \cite{Hassan-Rosen-bimetric} to add
a kinetic term for the metric $f_{\mu\nu}$ to the action and
converted it to an action for a bimetric massive theory which is
found to be ghost free. In the present work we generalize the work
in \cite{Hassan-Rosen-bimetric} to become a non-linear ghost free
action describing a trimetric model. The action then reads
\begin{align}\label{eq204}
S=&-M^2_g\int d^4x \sqrt{-g} R_g - M^2_f \int d^4 x \sqrt{-f} R_f - M^2_h \int d^4 x \sqrt{-h} R_h \nonumber\\
&+2m_1^2 M^2_{gf} \int d^4 x \sqrt{-g} \sum_{n=0}^{4} \beta _n e_n(\sqrt{g^{-1}f}) \nonumber\\
&+ 2m_2^{2} M^{2}_{gh} \int d^4 x \sqrt{-g} \sum_{m=0}^{4} \gamma _m e_m(\sqrt{g^{-1}h})\nonumber\\
&+2m_3^{ 2} M^{2}_{fh} \int d^4 x \sqrt{-f} \sum_{s=0}^{4} \alpha _s
e_s(\sqrt{f^{-1}h})
\end{align}
where $R_g$, $R_f$ and $R_h$ are the Ricci scalars constructed by three
metrics $g_{\mu\nu}$, $f_{\mu\nu}$ and $h_{\mu\nu}$ respectively. We
also have introduced three different Planck masses for each metric.
We have for each interaction term, an effective Planck mass
constructed by metrics included in the interaction
\begin{eqnarray}\label{eq205}
\frac{1}{M^{2}_{gf}}&\equiv&\frac{1}{M^2_g}+\frac{1}{M^2_f}\nonumber\\
\frac{1}{M^{2}_{gh}}&\equiv&\frac{1}{M^2_g}+\frac{1}{M^2_h}\nonumber\\
\frac{1}{M^{2}_{fh}}&\equiv&\frac{1}{M^2_f}+\frac{1}{M^2_h}.
\end{eqnarray}

\section{Linear theory}\label{sec3}
As is well known, bimetric theories in general, describe one
massless and one massive spin-2 fields. In this section we will see
that the above trimetric theory describes one
massless spin-2 field and two massive spin-2 fields which in general
interact with each other. We assume for simplicity, that the
interaction terms are all given by the minimal model introduced in
\cite{Hassan-Rosen-bimetric,ros}
\begin{align}\label{eq301}
\beta_0=3,\quad \beta_1=-1,\quad \beta_2=0,\quad\beta_3=0,\quad\beta_4=1,
\end{align}
and also for $\gamma_n$ and $\alpha_n$. In this model the
interaction terms can be written in terms of trace and determinant
\begin{align}\label{eq302}
2m_1^2 M^2_{gf} \int d^4 x \sqrt{-g}
\bigg(3-tr\sqrt{g^{-1}f}+det\sqrt{g^{-1}f}\bigg),
\end{align}
and similarly for the other interactions.

Now we expand the metrics around the same fixed
background\footnote{Note that our definition of perturbations is
different with \cite{Hassan-Rosen-bimetric} due to a factor. It does
not affect the final results.}
\begin{align}\label{eq303}
g_{\mu\nu}=\bar{g}_{\mu\nu}+\tilde{g}_{\mu\nu},\qquad
f_{\mu\nu}=\bar{g}_{\mu\nu}+\tilde{f}_{\mu\nu}, \qquad
h_{\mu\nu}=\bar{g}_{\mu\nu}+\tilde{h}_{\mu\nu}.
\end{align}
To second order in perturbations, this trimetric minimal model
reduces to the action
\begin{align}\label{eq304}
S=-\int
d^4x\bigg(&M_g^2\tg_{\mu\nu}\mc{E}^{\mu\nu\alpha\beta}\tg_{\alpha\beta}+M_f^2\tf_{\mu\nu}\mc{E}^{\mu\nu\alpha\beta}\tf_{\alpha\beta}+M_h^2
\th_{\mu\nu}\mc{E}^{\mu\nu\alpha\beta}\th_{\alpha\beta}\bigg)\nonumber\\
&-\f{1}{4}m_1^2 M_{gf}^2\int
d^4x\left[\left(\tg^\mu_{~\nu}-\tf^\mu_{~\nu}\right)^2-\left(\tg^\mu_{~\mu}-\tf^\mu_{~\mu}\right)^2\right]\nonumber\\
&-\f{1}{4}m_2^2 M_{gh}^2\int
d^4x\left[\left(\tg^\mu_{~\nu}-\th^\mu_{~\nu}\right)^2-\left(\tg^\mu_{~\mu}-\th^\mu_{~\mu}\right)^2\right]\nonumber\\
&-\f{1}{4}m_3^2 M_{fh}^2\int
d^4x\left[\left(\tf^\mu_{~\nu}-\th^\mu_{~\nu}\right)^2-\left(\tf^\mu_{~\mu}-\th^\mu_{~\mu}\right)^2\right],
\end{align}
where $\mc{E}^{\mu\nu\alpha\beta}$ is the Einstein-Hilbert kinetic term
and the background metric $\bar{g}_{\mu\nu}$ is responsible to raise
and lower the indices. Now by using the following transformations
\begin{eqnarray}\label{variable-transformation1}
\Upsilon'_{\mu\nu}&\equiv&\frac{1}{M^2}\left(M_g^2\tilde{g}_{\mu\nu}+M_f^2\tilde{f}_{\mu\nu}+M_h^2\tilde{h}_{\mu\nu}\right)\\
\Phi_{\mu\nu}&\equiv&\tilde{g}_{\mu\nu}-\tilde{f}_{\mu\nu}\\
\Psi_{\mu\nu}&\equiv&\tilde{f}_{\mu\nu}-\tilde{h}_{\mu\nu}\\
\Omega_{\mu\nu}&\equiv&\tilde{g}_{\mu\nu}-\tilde{h}_{\mu\nu}
\end{eqnarray}
where $M^2\equiv M_g^2+M_f^2+M_h^2$ and obviously
$\Phi_{\mu\nu}+\Psi_{\mu\nu}+\Omega_{\mu\nu}=0$, one can transform
the kinetic term in (\ref{eq304}) to
\begin{eqnarray}\label{kineticterm}
M^2\Upsilon'_{\mu\nu}\mc{E}^{\mu\nu\alpha\beta}\Upsilon'_{\alpha\beta}+\frac{M_g^2M_f^2M_h^2}{M^2M^2_{fh}}\Phi_{\mu\nu}\mc{E}^{\mu\nu\alpha\beta}\Phi_{\alpha\beta}+
\frac{M_g^2M_f^2M_h^2}{M^2M^2_{fg}}\Psi_{\mu\nu}\mc{E}^{\mu\nu\alpha\beta}\Psi_{\alpha\beta}
+\frac{M_g^2M_h^2}{M^2}\left[\Phi_{\mu\nu}\mc{E}^{\mu\nu\alpha\beta}\Psi_{\alpha\beta}+\Psi_{\mu\nu}\mc{E}^{\mu\nu\alpha\beta}\Phi_{\alpha\beta}\right]\nonumber
\end{eqnarray}
which is not diagonal due to the last two terms. Before making it
diagonal let us use
\begin{eqnarray}\label{variables-transformation1-2}\nonumber
\Phi'_{\mu\nu}\equiv
\frac{M_gM_fM_h}{MM_{fh}}\Phi_{\mu\nu},\hspace{1cm}
\Psi'_{\mu\nu}\equiv \frac{M_gM_fM_h}{MM_{gf}}\Psi_{\mu\nu}\nonumber
\end{eqnarray}
To make them diagonal one may now use the following
transformations
\begin{eqnarray}\label{variables-transformation2}\nonumber
\Phi''_{\mu\nu}\equiv\frac{1}{2}\left(\Phi'_{\mu\nu}+\Psi'_{\mu\nu}\right),\hspace{1cm}
\Psi''_{\mu\nu}\equiv\frac{1}{2}\left(\Phi'_{\mu\nu}-\Psi'_{\mu\nu}\right)\nonumber,
\end{eqnarray}
to get
\begin{eqnarray}\label{kineticterm2}
M^2\Upsilon'_{\mu\nu}\mc{E}^{\mu\nu\alpha\beta}\Upsilon'_{\alpha\beta}+
2\left(1+\frac{M_{fh}M_{gf}}{M_f^2}\right)\Phi''_{\mu\nu}\mc{E}^{\mu\nu\alpha\beta}\Phi''_{\alpha\beta}+
2\left(1-\frac{M_{fh}M_{gf}}{M_f^2}\right)\Psi''_{\mu\nu}\mc{E}^{\mu\nu\alpha\beta}\Psi''_{\alpha\beta}\nonumber,
\end{eqnarray}
which is not yet canonical due to different weights. Eventually by
assuming
\begin{eqnarray}\label{variables-transformation3}\nonumber
\Upsilon_{\mu\nu}\equiv M\Upsilon'_{\mu\nu},\hspace{.5cm}
\Pi_{\mu\nu}\equiv\sqrt{2\left(1+\frac{M_{fh}M_{gf}}{M_f^2}\right)}\Phi''_{\mu\nu},\hspace{.5cm}
\Xi_{\mu\nu}\equiv\sqrt{2\left(1-\frac{M_{fh}M_{gf}}{M_f^2}\right)}\Psi''_{\mu\nu}\nonumber,
\end{eqnarray}
the kinetic term will be canonical as follows
\begin{eqnarray}\label{kineticterm3}
\Upsilon_{\mu\nu}\mc{E}^{\mu\nu\alpha\beta}\Upsilon_{\alpha\beta}+\Pi_{\mu\nu}\mc{E}^{\mu\nu\alpha\beta}\Pi_{\alpha\beta}+
\Xi_{\mu\nu}\mc{E}^{\mu\nu\alpha\beta}\Xi_{\alpha\beta}\nonumber,
\end{eqnarray}
with new fields $\Upsilon$, $\Pi$ and $\Xi$. Let us now use the above
procedure for the potential term in  Lagrangian (\ref{eq304}).
The corresponding potential term with a straightforward algebraic
calculation is
\begin{eqnarray}\label{potentialterm1}
-\frac{M^2_1}{4}\left(\Pi_{\mu\nu}\Pi^{\mu\nu}-\Pi^\mu_{\hspace{1.5mm}\mu}\Pi^\nu_{\hspace{1.5mm}\nu}\right)-
\frac{M^2_2}{4}\left(\Xi_{\mu\nu}\Xi^{\mu\nu}-\Xi^\mu_{\hspace{1.5mm}\mu}\Xi^\nu_{\hspace{1.5mm}\nu}\right)-
\frac{\lambda}{4}\left(\Pi_{\mu\nu}\Xi^{\mu\nu}-\Pi^\mu_{\hspace{1.5mm}\mu}\Xi^\nu_{\hspace{1.5mm}\nu}\right)\nonumber,
\end{eqnarray}
where
\begin{eqnarray}\label{new-coefs}\nonumber
M^2_1&\equiv&\frac{1}{2}\left(1+\frac{M_{fh}M_{gf}}{M_f^2}\right)^{-1}\frac{M^2}{M_g^2M_f^2M_h^2}\bigg[M^2_{fh}M_{gf}^2(m_1^2+m_3^2)+(M_{fh}+M_{gf})^2(m_2^2M_{gh}^2)\bigg]\\\nonumber
M^2_2&\equiv&\frac{1}{2}\left(1-\frac{M_{fh}M_{gf}}{M_f^2}\right)^{-1}\frac{M^2}{M_g^2M_f^2M_h^2}\bigg[M^2_{fh}M_{gf}^2(m_1^2+m_3^2)+(M_{fh}-M_{gf})^2(m_2^2M_{gh}^2)\bigg]\\\nonumber
\lambda&\equiv&\left(1-\frac{M^2_{fh}M^2_{fg}}{M_f^4}\right)^{-\frac{1}{2}}\frac{M^2}{M_g^2M_f^2M_h^2}\bigg[M^2_{fh}M_{gf}^2(m_1^2-m_3^2)+m_2^2M_{gh}^2(
M_{fh}^2-M_{gf}^2)\bigg].
\end{eqnarray}
The full Lagrangian for these new variables is therefore
\begin{eqnarray}\label{lagrangian}
{\cal{L}}=&&\bigg(\Upsilon_{\mu\nu}\mc{E}^{\mu\nu\alpha\beta}\Upsilon_{\alpha\beta}+\Pi_{\mu\nu}\mc{E}^{\mu\nu\alpha\beta}\Pi_{\alpha\beta}+
\Xi_{\mu\nu}\mc{E}^{\mu\nu\alpha\beta}\Xi_{\alpha\beta}\bigg)\\\nonumber-&&\bigg[\frac{M^2_1}{4}\left(\Pi_{\mu\nu}\Pi^{\mu\nu}-\Pi^\mu_{\hspace{1.5mm}\mu}\Pi^\nu_{\hspace{1.5mm}\nu}\right)+
\frac{M^2_2}{4}\left(\Xi_{\mu\nu}\Xi^{\mu\nu}-\Xi^\mu_{\hspace{1.5mm}\mu}\Xi^\nu_{\hspace{1.5mm}\nu}\right)+
\frac{\lambda}{4}\left(\Pi_{\mu\nu}\Xi^{\mu\nu}-\Pi^\mu_{\hspace{1.5mm}\mu}\Xi^\nu_{\hspace{1.5mm}\nu}\right)\bigg].\nonumber
\end{eqnarray}
It is obvious that one of the gravitons, $\Upsilon$, is massless
though the other gravitons not only are massive but also interactive
with arbitrary coefficients. This interaction term is a new
prediction for having more than two gravitons and cannot be resolved
by a new redefinition of the fields while taking the kinetic term
canonical. This fact can be emphasized by noting that in the primary
Lagrangian (\ref{eq304}), three dimensional parameters exist.
Consequently, these three dimensional parameters appear
naturally after the field redefinitions. However, now two of
them are represented as mass terms and the other as an
interaction term. However, it is possible to resolve the interaction term
by  fine-tuning as ``$m^2_1=m^2_3$ and $M_{h}^2=M_{g}^2$" which
means that in the primary Lagrangian (\ref{eq304}) there are just two
dimensional parameters. An interesting point is that the massless
graviton  corresponds to
$$\Upsilon_{\mu\nu}\propto\left(M_g^2\tilde{g}_{\mu\nu}+M_f^2\tilde{f}_{\mu\nu}+M_h^2\tilde{h}_{\mu\nu}\right),$$
which is exactly the average of the primary fields in (\ref{eq304}).
To have an idea on the masses, let us assume the special case of
$m^2=m_1^2=m_2^2=m_3^2$ and $\sqrt{3}M=M_g=M_f=M_h$ which means we
have just one dimensional scale with the primary Lagrangian
(\ref{eq204}) being totally symmetric under changing the metrics. In
this case the final gravitons have masses $0$ and
$M_1^2=M_2^2=3\times\frac{m^2}{2}$ where $\frac{m^2}{2}$ appears in (\ref{eq304})
as mass term\footnote{Note that the dividing by ``$2$" is a consequence of our
assumptions for this special case i.e. $m^2=m_1^2=m_2^2=m_3^2$ and $\sqrt{3}M=M_g=M_f=M_h$
and definitions (\ref{eq205})}. The interesting point is that the coefficient ``$3$" is exactly
the number of metrics which exist in the model. This argument can be generalized for $N$-metric
formalism by a little linear algebra for matrices. So in $N$-metric formalism,
one graviton is massless and other $N-1$ gravitons have masses as $M_i^2=N\times m^2,\forall i \in \{1,2,...,N-1\}$
where $m^2$ is the mass before the field redefinitions.

It should be noticed that some of the above conclusion is true for the
$N$-metric formalism. In fact in $N$-metric models with  ghost-free
potential terms such as those in (\ref{eq304}), after the redefinition of
fields, the model represents a massless graviton which is the
average of the metrics and the $N-1$ massive gravitons with
interaction. For a quick check let us innumerate the number of
dimensional parameters. For the primary Lagrangian the number of interaction
terms are $\big(^N_{\hspace{.5mm}2}\big)$, but after field
redefinition one field becomes massless. The other $N-1$ remaining fields
are massive and present $N-1$ dimensional parameters as well as
$\big(^{N-1}_{\hspace{2.2mm}2}\big)$ new interactive terms. So
finally we have $N-1+\big(^{N-1}_{\hspace{2.2mm}2}\big)$
dimensional parameters which is exactly
$\big(^N_{\hspace{.5mm}2}\big)$. In the language of
\cite{nima-a-h-1}, this model can be understood as follows; the
primary Lagrangian (\ref{eq204}) represents $N$ sites which are
linked to each other with $\big(^N_{\hspace{.5mm}2}\big)$ links.
This representation can be reduced to a representation with $N$
sites with $\it{one}$ site which has no link to the remaining
$N-1$ sites. But these $N-1$ sites are linked not only to each other
but also have self-links or respectively have interactions and
masses.

\subsection{Coupling to matter}
As mentioned in \cite{Hassan-Rosen-bimetric}, the coupling of
multi-metric theories in the context of massive gravity to
matter fields has not yet been solved. However, by considering
the symmetry it seems as if one can constrain the coupling of
the matter to gravitons. By looking at (\ref{eq304}) and assuming that
there is no preference between the primary metrics one may then
conclude that each kind of coupling is natural to be symmetric under
the interchange of the metrics. In addition, from the linearized
theory, it is obvious that  there is a natural choice for the
symmetric form of the metric which can be constructed out of their average. So,
just by considering the symmetry, it seems that any correct coupling
of matter should be to a symmetric combination of metrics which
reduces to their average in the linear form. This argument becomes
more significant if we note that the average of the metrics in
the linear theory is simply the massless graviton. This means that whenever
the linear theory is applicable, the matter couples to massless gravitons
directly and does not see the massive ones. The existence of
massive gravitons may show themselves in higher order terms which in turn may
be verified in the context of cosmological perturbations.
In the power spectrum of cosmological perturbations (e.g. curvature
perturbation) there is no difference between Einstein gravity and
massive gravity. However, in the bispectrum of perturbations (i.e.
non-Gaussianity) the massive gravity should have different
predictions.

\section{Cosmological solutions}\label{sec4}
\subsection{Cosmological equations}
In this section we obtain the cosmological solutions of the above
trimetric theory. In the massive gravity case, it has been shown
that the flat FRW space-time is not the solution of the model
\cite{cos}. However, open FRW solutions do exist \cite{muko}.
Recently, the cosmological solutions for the non-linear ghost-free
bimetric action was obtained in \cite{volkov,hassan}. Now we generalize
them to the trimetric case.

Let us assume that all three metrics can be described in terms of an
isotropic and homogeneous space-times
\begin{align}\label{eq401}
ds_g^2&=-N(t)^2dt^2+a(t)^2\left(dr^2+r^2 d\Omega^2\right)\nonumber\\
ds_f^2&=-M(t)^2dt^2+b(t)^2\left(dr^2+r^2 d\Omega^2\right)\nonumber\\
ds_h^2&=-Q(t)^2dt^2+w(t)^2\left(dr^2+r^2 d\Omega^2\right).
\end{align}
Plugging these into  action \eqref{eq204}, we can read the reduced
Lagrangian as follows
\begin{align}\label{eq402}
\mathcal{L}_{red}&=6\left(M_g^2\f{a\dot{a}^2}{N}+M_f^2\f{b\dot{b}^2}{M}+M_h^2\f{w\dot{w}^2}{Q}\right)\\
&+2m_1^2M_{gf}^2\bigg(\beta_0Na^3+\beta_1(Ma^3+3Na^2b)+3\beta_2ab(Ma+Nb)+\beta_3b^2(3Ma+Nb)
+\beta_4Mb^3\bigg) \nonumber\\
&+2m_2^2M_{gh}^{2}\bigg(\gamma_0Na^3+
\gamma_1(Qa^3+3Na^2w)+3\gamma_2 aw(Qa+Nw)+\gamma_3 w^2(3Qa+Nw)
+\gamma_4 Qw^3\bigg) \nonumber\\\nonumber
&+2m_3^2M_{fh}^{2}\bigg(\alpha_0 Mb^3+
\alpha_1(Qb^3+3Mb^2w)+3\alpha_2 bw(Qb+Mw)+\alpha_3 w^2(3Qb+Mw)
+\alpha_4 Qw^3\bigg).
\end{align}
The equations of motion can be obtained by varying the reduced
Lagrangian with respect to scale factors with the result
\begin{align}\label{eq403}
6M_g^2&\left(\f{\dot{a}^2}{N}+2\f{a\ddot{a}}{N}-2\f{a\dot{a}\dot{N}}{N^2}\right)\nonumber\\
&-2m_1^2M_{gf}^2\bigg(3\beta_0Na^2+3\beta_1a(Ma+2Nb)+3\beta_2b(2Ma+Nb)+3\beta_3Mb^2\bigg)\nonumber\\
&-2m_2^2M_{gh}^{2}\bigg(3\gamma_0Na^2+3\gamma_1a(Qa+2Nw)+3\gamma_2w(2Qa+Nw)+3\gamma_3Qw^2\bigg)=0,
\end{align}
\begin{align}\label{eq404}
6M_f^2&\left(\f{\dot{b}^2}{M}+2\f{b\ddot{b}}{M}-2\f{b\dot{b}\dot{M}}{M^2}\right)\nonumber\\
&-2m_1^2M_{gf}^{2}\bigg(3\beta_1Na^2+3\beta_2a(Ma+2Nb)+3\beta_3b(2Ma+Nb)+3\beta_4Mb^2\bigg)\nonumber\\
&-2m_3^2M_{fh}^{2}\bigg(3\alpha_0Mb^2+3\alpha_1b(Qb+2Mw)+3\alpha_2w(2Qb+Mw)+3\alpha_3Qw^2\bigg)=0,
\end{align}
\begin{align}\label{eq405}
6M_h^2&\left(\f{\dot{w}^2}{Q}+2\f{w\ddot{w}}{Q}-2\f{w\dot{w}\dot{Q}}{Q^2}\right)\nonumber\\
&-2m_2^2M_{gh}^{2}\bigg(3\gamma_1Na^2+3\gamma_2a(Qa+2Nw)+3\gamma_3w(2Qa+Nw)+3\gamma_4Qw^2\bigg)\nonumber\\
&-2m_3^2M_{fh}^{2}\bigg(3\alpha_1Mb^2+3\alpha_2b(Qb+2Mw)+3\alpha_3w(2Qb+Mw)+3\alpha_4Qw^2\bigg)=0,
\end{align}
and variation of the Lagrangian with respect to the lapse functions become
\begin{align}\label{eq406}
6M_g^2\f{a\dot{a}^2}{N^2}&-2m_1^2M_{gf}^2\bigg(\beta_0a^3+3\beta_1ba^2+3\beta_2ab^2+\beta_3b^3\bigg)\nonumber\\
&-2m_2^2M_{gh}^{2}\bigg(\gamma_0a^3+3\gamma_1wa^2+3\gamma_2aw^2+\gamma_3w^3\bigg)=0,
\end{align}
\begin{align}\label{eq407}
6M_f^2\f{b\dot{b}^2}{M^2}&-2m_1^2M_{gf}^2\bigg(\beta_1a^3+3\beta_2ba^2+3\beta_3ab^2+\beta_4b^3\bigg)\nonumber\\
&-2m_3^2M_{fh}^{2}\bigg(\alpha_0b^3+3\alpha_1wb^2+3\alpha_2bw^2+\alpha_3w^3\bigg)=0,
\end{align}
\begin{align}\label{eq408}
6M_h^2\f{w\dot{w}^2}{Q^2} &-2m_2^2M_{gh}^{2}\bigg(\gamma_1a^3+3\gamma_2wa^2+3\gamma_3aw^2+\gamma_4w^3\bigg)\nonumber\\
&-2m_3^2M_{fh}^{2}\bigg(\alpha_1b^3+3\alpha_2wb^2+3\alpha_3bw^2+\alpha_4w^3\bigg)=0.
\end{align}

\subsection{Solution I}
In this section we are going to show that there is a solution for
the above equations. To do this we suppose
\begin{eqnarray}
a=b=w
\end{eqnarray}
and $N=M=Q=1$, that is, we work in the comoving gauge. Equation (\ref{eq406})
reduces to two disjoint \footnote{The linear combination of their
solutions is not a solution.} equations
\begin{eqnarray}\label{eq1-a=b}
\dot{a}&=&\pm\omega a,\\\nonumber
\omega&\equiv&\frac{1}{\sqrt{3}M_g}\bigg[m_1^2M^2_{gf}\bigg(\beta_0+3\beta_1+3\beta_2+\beta_3\bigg)+
m_2^2M^2_{gh}\bigg(\gamma_0+3\gamma_1+3\gamma_2+\gamma_3\bigg)\bigg]^{\frac{1}{2}}.\\\nonumber
\end{eqnarray}
Obviously, the solutions are $a=e^{-\omega t}$ and $a=e^{+\omega t}$
for minus and plus signs respectively. The latter, which is more
interesting, is a de-Sitter solution with a cosmological constant
$\Lambda=3 \omega^2$. Now it should be checked if this
solution also satisfies the first equation in (\ref{eq403}) for $a=
b$ in the comoving gauge. This equation is
\begin{eqnarray}\label{eq2-a=b}
\dot{a}^2+2a\ddot{a}=3 \omega^2a^2,
\end{eqnarray}
where $\omega$ has been defined previously. Obviously, both of
$a=e^{-\omega t}$ and $a=e^{+\omega t}$ satisfy the above equation.
Now we should check what are the conditions imposed on the remaining
equations. One can simply find that the following equation should be
satisfied due to (\ref{eq407}) and (\ref{eq408})
\begin{eqnarray}\label{constraint-parameters}\nonumber
\omega&\equiv&\frac{1}{\sqrt{3}M_g}\bigg[m_1^2M^2_{gf}\bigg(\beta_0+3\beta_1+3\beta_2+\beta_3\bigg)+
m_2^2M^2_{gh}\bigg(\gamma_0+3\gamma_1+3\gamma_2+\gamma_3\bigg)\bigg]^{\frac{1}{2}}\\\nonumber
&=&\frac{1}{\sqrt{3}M_f}\bigg[m_1^2M_{gf}^2\bigg(\beta_1+3\beta_2+3\beta_3+\beta_4\bigg)
+m_3^2M_{fh}^{2}\bigg(\alpha_0+3\alpha_1+3\alpha_2+\alpha_3\bigg)\bigg]^{\frac{1}{2}}\\\nonumber
&=&\frac{1}{\sqrt{3}M_h}\bigg[m_2^2M_{gh}^{2}\bigg(\gamma_1+3\gamma_2+3\gamma_3+\gamma_4\bigg)
+m_3^2M_{fh}^{2}\bigg(\alpha_1+3\alpha_2+3\alpha_3+\alpha_4\bigg)\bigg]^{\frac{1}{2}},
\end{eqnarray}
where the first line is the definition of $\omega$ and the last
lines should be seen as constraints on the parameters such that
$a=b=w=e^{\pm \omega t}$. It is obvious that equations (\ref{eq404})
and (\ref{eq405}) are  automatically satisfied  as equation
(\ref{eq403}).

\subsection{Solution II}
Let us now see if in such a massive cosmology with
more than two metrics it is possible to have a solution with one
Minkowski metric. This solution is impossible for the bimetric
case\footnote{This is obvious from recent works on this topic
\cite{volkov} and \cite{hassan} though they did not mention it. The
procedure is similar to what we have done for trimetric model.}. This
assumption may be interesting because of its resemblance to
ordinary massive cosmology \cite{cos} where a static metric together with a
dynamical one exist.

Using equation (\ref{eq408}) and assuming that $w$ is a constant, one
gets the following relation
\begin{align}\label{relation-between-a-b}\nonumber
m_2^2M_{gh}^{2}\bigg(\gamma_1a^3+3\gamma_2wa^2+3\gamma_3aw^2+\gamma_4w^3\bigg)
=-m_3^2M_{fh}^{2}\bigg(\alpha_1b^3+3\alpha_2wb^2+3\alpha_3bw^2+\alpha_4w^3\bigg)
\end{align}
which should be satisfied for all $a$ and $b$. Since $a$ and $b$ are
dynamical fields, the above condition is satisfied if and only if
RHS and LHS are equal term by term. The general solution for this
is as follows \footnote{One can assume $a=\zeta b$ and $w=c$
for constant $\zeta$ and $c$. But these assumptions will reduce to
(\ref{a-b-2}) by a re-definition of $\alpha_i$ and $\gamma_i$
$\forall i\in \{1,2,3,4\}$ without any physical consequences.}
\begin{eqnarray}\label{a-b-2}
a=b,\hspace{1cm}w=1,\hspace{1cm}m_2^2M_{gh}^{2}\gamma_i=-m_3^2M_{fh}^{2}\alpha_i\hspace{.5cm}
i\in\{1,2,3,4\}.
\end{eqnarray}
What is done here represents a possible
cosmological solution for the trimetric formalism of massive gravity.
This solution is deduced by assuming that one of the metrics is
static. This property makes this solution interesting and comparable
to the assumptions in \cite{cos}.
Now, equation (\ref{eq406}) with the above assumption becomes
\begin{eqnarray}\label{eq-a-a=b}
6M_g^2a\dot{a}^2=2m_1^2M_{gf}^2\bigg(\beta_0+3\beta_1+3\beta_2+\beta_3\bigg)a^3+2m_2^2M_{gh}^{2}\bigg(\gamma_0a^3+3\gamma_1a^2+3\gamma_2a+\gamma_3\bigg),
\end{eqnarray}
and equation (\ref{eq407}) with $a=b$
\begin{eqnarray}\label{eq-b-a=b}
6M_f^2a\dot{a}^2=2m_1^2M_{gf}^2\bigg(\beta_1+3\beta_2+3\beta_3+\beta_4\bigg)a^3
+2m_3^2M_{fh}^{2}\bigg(\alpha_0a^3+3\alpha_1a^2+3\alpha_2a+\alpha_3\bigg).
\end{eqnarray}
By comparing the above two equations and our assumption of taking one
of the metrics as static (\ref{a-b-2}), it is easy to see that the
following conditions have to be satisfied by the parameters
\begin{eqnarray}\nonumber
\frac{1}{M_g^2}\bigg[m_1^2M_{gf}^2\big(\beta_0+3\beta_1+3\beta_2+\beta_3\big)+m_2^2M_{gh}^{2}\gamma_0\bigg]&=&
\frac{1}{M_f^2}\bigg[m_1^2M_{gf}^2\big(\beta_1+3\beta_2+3\beta_3+\beta_4\big)+m_3^2M_{fh}^{2}\alpha_0\bigg]\\\nonumber\gamma_i=\alpha_i=0,&&\hspace{.5cm}
i\in\{1,2,3\}.
\end{eqnarray}
Actually, with the above conditions one should consider the
remaining equations to get a correct solution. However, there is no need to pursue the matter any further in this case since, as we saw, the necessary condition of taking
one of the metrics static is the same as assuming $\gamma_i=\alpha_i=0$ for
$i\in\{1,2,3\}$. By looking  at  Lagrangian (\ref{eq402}), it is
easy to find that this condition prevents
the third metric in (\ref{eq401}) interacting with others. So this solution
is the trivial one and is without any importance.

Finally, we showed  that as a result of
\cite{cos}, there is no non-trivial solution for the massive
cosmology studied here if, at least, one of the metrics is static.

\section{Conclusions and Questions}\label{conclu}
We have considered the generalization of the work done in
\cite{Hassan-Rosen-bimetric}, from a bimetric theory to a trimetric theory.
This trimetric theory in fact has many of the
consequences of the $N$-metric theory, with the advantage of
simplicity. What we get in this paper, is the conclusion that, in
the multi-metric theories, only one graviton becomes massless at the
linear level, which is the average of the $N$ metrics. The other
$N-1$ spin-2 fields become massive and in general interact with each
other. This has a general conclusion that  ordinary matter must
couple to  metrics in such a way that at the linear level only
the coupling to the average metric survives. This constraint on the
form of the metric may be important when one addresses
the subtleties of the multi-metric formalism.
In this formalism, the geometrical interpretation of metrics e.g.
the meaning of the distance and covariant derivative (parallel transportation)
are yet unresolved. But with the above restriction which comes from coupling to matter
a large class of  multi-metric formulations are ruled out.

The cosmology of this multi-metric theory is also interesting and generalizes the result in
massive cosmology \cite{cos}. We have seen in this paper that, if we take one of the metrics as
Minkowskian, it will become impossible to get a
non-trivial cosmological solution for the theory. This result is
however true only in a flat case. It is a matter of further investigation to see if
the theory can have  non-trivial non-flat cosmological solutions,
as in the massive cosmology case \cite{muko}. It is also crucial
to consider the cosmological perturbation for FRW cosmology in this
context due to the existence of a lot of observational data. From a theoretical viewpoint,
when having more than one metric, it is natural to get not only
the adiabatic mode but also entropy perturbations without any need for
extra matter fields.

We should also mention that the multi-metric theory
constructed here is not proven to be ghost-free \footnote{However it
seems the mechanism for bimetric model (see the second reference in
\cite{Hassan-Rosen-bimetric}) may be applicable.}. In fact, it is
possible that the interaction terms introduced here may not
suffice to cancel all Boulware-Deser ghosts of all spin-2
fields. This would require further investigation in a future work.

\begin{acknowledgments}
We would like to thank B. Bassett, P. Creminelli, C. de Rham and P.
Moyassari for their comments and discussions. N. K. would like to
thank P. Creminelli for very fruitful discussions during his stay at
ICTP and the organizers of the Workshop on Infrared Modifications of
Gravity.
\end{acknowledgments}


\end{document}